# Water skin anomalies: density, elasticity, hydrophobicity, thermal stability, interface repulsivity, etc

Chang Q Sun

Ecqsun@ntu.edu.sg

Nanyang Technological University, Singapore

**Molecular undercoordination induced O:H-O bond relaxation and dual polarization dictates the supersolid behavior of water skins interacting with other substances such as flowing in nanochannels, dancing of water droplets, floating of insects. The BOLS-NEP notion unifies the Wenzel-Cassie-Baxter models and explains controllable transition between hydrophobicity and hydrophilicity.**

**Contents**

## 1   Anomalies: Skin matters

Water skin demonstrates numerous anomalies, as epitomized in Figure 1 and Figure 2:

1) Water skin has the highest tension of ever known and the tension drops when heated.
2) Water skin is elastic, hydrophobic, less dense, tough, and thermally more stable, which is more pronounced at elevated capillary curvature or hydrophobically confined.

Water on water, water on certain kinds of substance or inverse shows the same attributes with however unclear mechanism. Amazingly, small insects such as a strider can stand, walk and glide on water freely. A water strider statically standing on water can bear a load up to ten times its body weight with its middle and hind legs, which tread deep puddles without piercing the water skin [1]. If carefully placed on the skin, a small needle, or a coin, floats on water even though its density is times higher than that of water because: (i) it weighs insufficiently to penetrate the skin and (ii) the interface between its paddle and the skin of water is hydrophobic. If the surface is agitated to break up the tension, then the needle will sink quickly. The extraordinary hydrophobicity and toughness of water skin are attributed to the presence of a layer of molecules in solid state [2, 3]. Whether a solid skin forms on water or a liquid skin covers ice has long been a paradox.

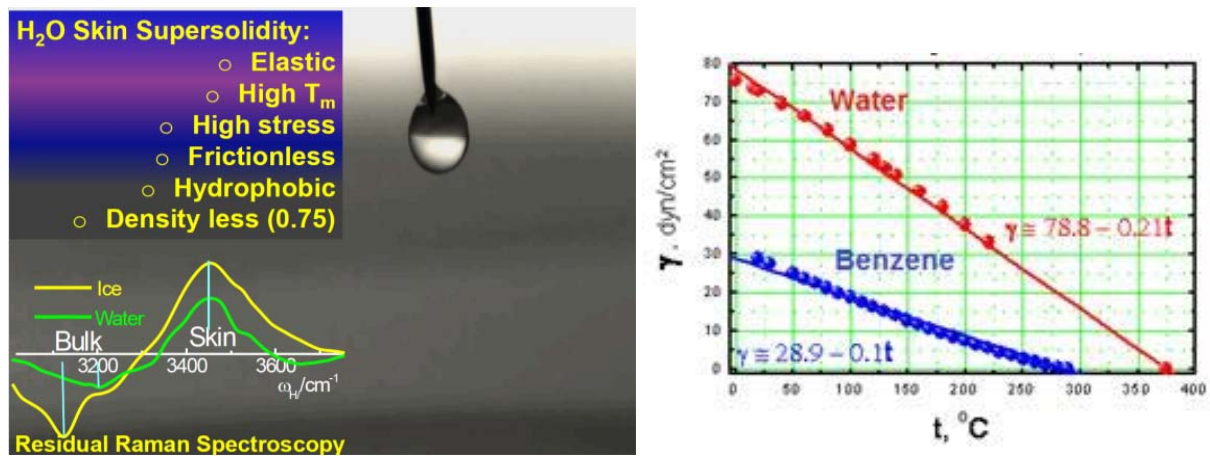

Figure 1. Does ice cover water or water cover ice? (a) A video clip [4] shows that a water droplet bounces continuously and repeatedly on water, which evidences the elasticity and hydrophobicity of both skins of the bulk water and the droplet. (b) Surface tension of water and benzene at different temperatures.

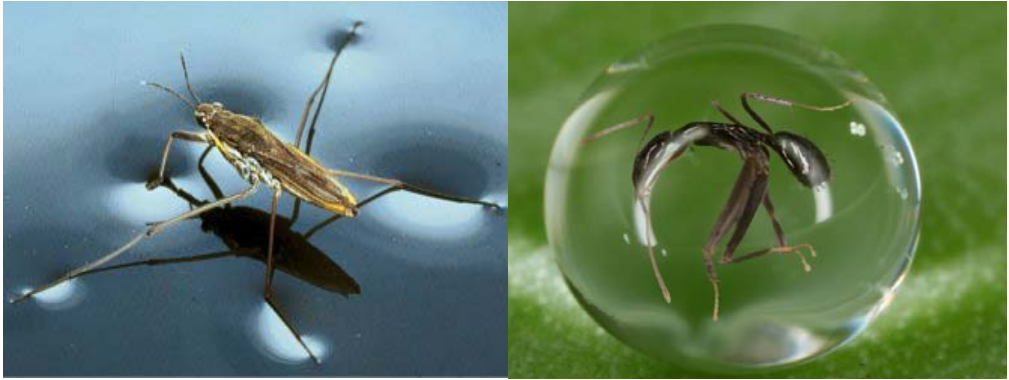

Figure 2. (a) A strider can stand still and slide on water and (b) an ant trapped in a tiny perfect sphere of water, totally unable to escape, after being caught in a sudden heavy downpour (Picture: Adam Gormley, Queensland, Australia, 2011)[5].

## 2  *Reasons: Skin supersolidity*

From the perspective of molecular undercoordination and the O:H-O bond inter-oxygen repulsion [6], the following explains anomalies of water skin (also refer to sections dealing with ice and water clusters), see Figure 3:

1) Undercoordinated water molecules shrink their sizes (H-O bond length) and expand their separations (O:H nonbond length) through inter-oxygen repulsion and dual polarization; H-O contraction and O:H expansion result in density loss by up to 25%.
2) H-O bond stiffening raises H-O phonon frequency $\omega_H$, melting point $T_m$, O 1s energy shift $E_{1s}$ and $E_H$ for H-O atomic dissociation; O:H softening lowers O:H phonon frequency $\omega_L$, boiling temperature $T_V$, and $E_L$ for molecular dissociation with negligible contribution to the $E_{1s}$.
3) Shared by ice and water droplet, polarization raises the hydrophobicity, elasticity, and repulsivity, making skins supersolid characterized by an identical $\omega_H$ of 3450 cm$^{-1}$.
4) Undercoordinated skin molecules are subject to forces of 'skin repulsion' instead of 'surface tension'.

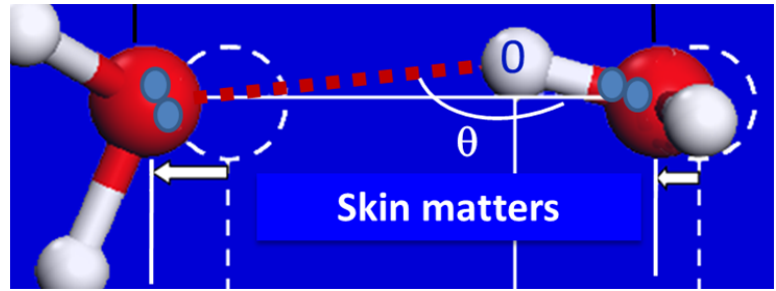

Figure 3. Molecular undercoordination shortens and stiffens the H-O bond and lengthens and softens the O:H nonbond associated with dual polarization result in the anomalous supersolid skin of water and ice [6].

## 3 History background

??

Skin tension helps seeds bury themselves by causing awns to coil and uncoil. It enables a floating fern to maintain an air layer, even when submerged. It also makes a beetle fly in two dimensions, not three. Surface tension also allows human and agricultural pathogens to travel long distances in tiny, buoyant droplets. The hardly noticed skin tension does play a big role in life at large [7].

## 4 Quantitative evidence

### 4.1 Elasticity and Hydrophobicity

Firstly, water skin is ice-like at ambient temperature. SFG spectral measurements and MD calculations suggested that the outermost two layers of water molecules have an 'ice-like' order at room temperature [8]. At ambient temperature, ultrathin films of water perform like ice with a hydrophobic nature [3, 9]. Water at temperatures of 7, 25, and 66°C at atmospheric pressure has an ordered skin 0.04–0.12 nm thick [10].

Secondly, air gap presents at hydrophobic contacts. Using specular x-ray reflectivity analysis, Uysal et al [11] suggested that an air gap of 0.5–1.0 nm thick exists between the water the hydrophobic substrate.

The air gap increases with the contact angle of the droplet curvature or with the lowering of the effective CN of molecules at the skin.

A video clip [4] (see Figure 1a) shows that a water droplet bounces continuously and repeatedly on water, which evidences straightforwardly the elasticity and hydrophobicity of water skin regardless of curvature. Consistency between theoretical calculations and measurements further confirmed that a monolayer film of water manifests 'quasi-solid' behavior at room temperature, and a hydrophobic nature that prevents it from being wetted itself by a water droplet [9, 12], see Figure 4.

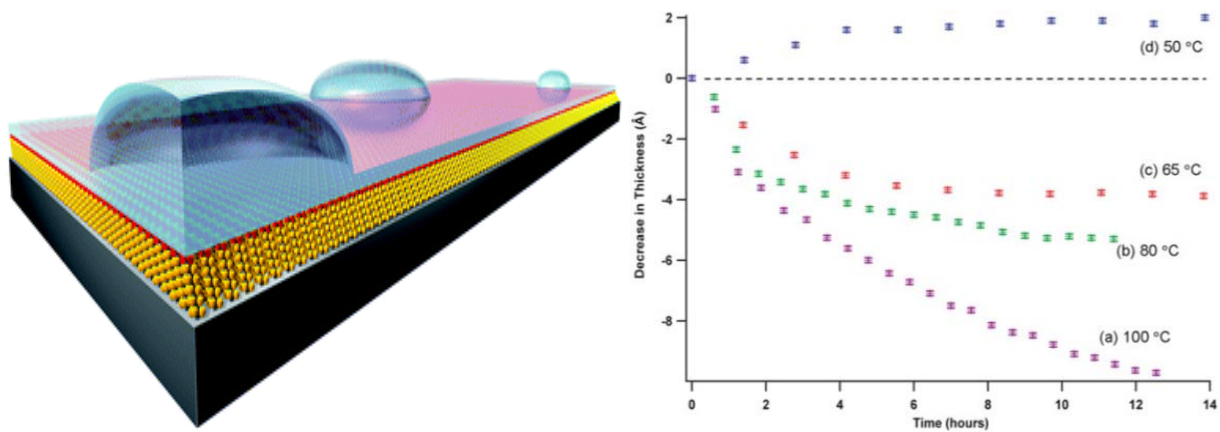

Figure 4. (a, not to scale) Nanoscaled water droplet in contact with a continuous water layer deposited on a hydrophilic COOH-terminated monolayer is "ice-like" at room temperature. (b) Time and temperature dependence of water films, indicating the hydrophobicity and thermal stability (H-O bond energy dictates melting temperature yet O:H nonbond energy determines evaporation) of ultrathin water layer at room temperature [9, 12]. The critical temperature stabilizes the film thickness is between 50 and 65° C and evaporation occurs around 65 °C means the lowered evaporation temperature. (Reprinted with permission from [12].)

Using X-ray and neutron reflectometry and atomic force microscopy (AFM), James and co-workers [12] demonstrated that water is almost universally present on apparently dry self-assembled monolayers, even on those considered almost hydrophobic by conventional methods such as water contact goniometry. They observed condensation of water on hydrophilic surfaces under ambient conditions formed a dense sub-nanometer surface layer; the thickness of which increased with exponentially limiting kinetics. Tapping mode AFM measurements show the presence of nanosized droplets that covered about

2% of the total surface area, and which became fewer in number and larger in size with time. While high vacuum (~$10^{-8}$ bar) at room temperature could hardly remove the adsorbed water droplet from these monolayers; heating to temperatures above 65 °C under atmospheric conditions results in evaporation from the surface.

It has been shown that the H-O bond energy dictates temperature of melting while the O:H nonbond energy determines evaporation. Findings of James and co-workers [12] confirmed the hydrophobicity and thermal stability of ultrathin water film at room-temperature predicted theoretically [9]. The melting point is higher and the evaporation point is lower compared with bulk water, being consistent with the present BOLS-NEP notion prediction.

Water droplets also dance on solid surfaces, regardless of substrate temperatures and materials (-79°C $CO_2$; 22°C superhydrophobic surface, and 300°C Al plate) [13]. These observations were attributed to the Leidenfrost effect, first reported in 1756 [14] and the superhydrophobicity [15] known from the late 1950s for room temperature and substrate sublimation effects. In the Leidenfrost condition on a hot substrate, the impacting liquid drop rapidly forms a vapor layer at the liquid-substrate interface. This vapor layer (with a thickness typically in the range of 10-100 μm) acts both as cushion and as thermal insulator, causing a freely floating and gradually evaporating drop.

On a superhydrophobic surface, the contact between the water drop and the solid substrate is only partial, due to the presence of a composite air-liquid-solid interface, where air pockets prevent full surface wetting. The wetting area is typically less than 20% of the total solid surface area. As a result, water drops move easily on the surface, due to low adhesion capillary forces between liquid drop and the solid substrate. The lower contact areas and air pockets at the interface result from the substrate atomic undercoordination effect. Atomic undercoordination shortens the bond causing local densification and entrapment of bonding and core electrons, which in turn polarize the nonbonding electrons pertaining to undercoordinated edge atoms, providing repulsive force and making substrate superhydrophobic.

Drop rebounds from the substrate at -79 °C temperatures, which indicates sublimation of the solid surface occurs by frost formation, preventing the surface from being contacted to the solid $CO_2$ skin [13].

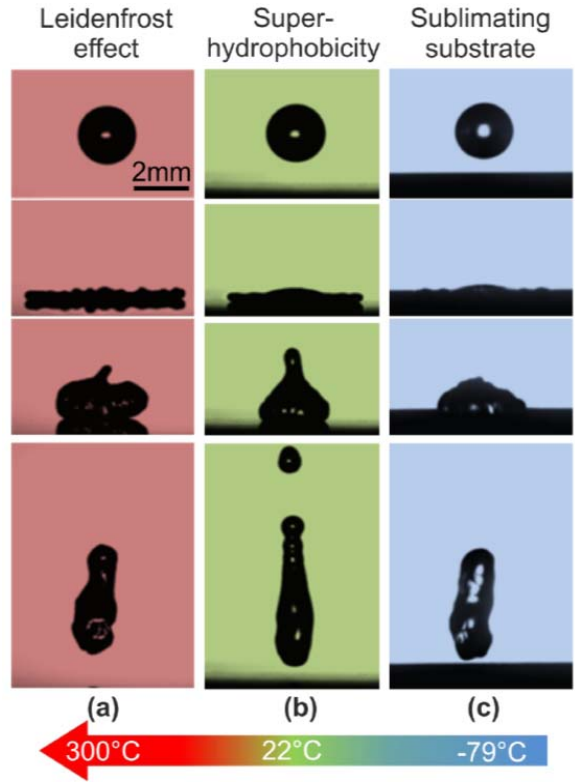

Figure 5. Water droplets impacting on (a) Al sample at 300 °C (subject to Leidenfrost effect – vapor formed below the droplet), (b) superhydrophobic surface (subject to superhydrophobicity – atomic undercoordination induced substrate quantum entrapment and polarization [16]), and (c) solid $CO_2$ at -79 °C (subject to sublimating – frost formation between contacts). (Reprinted with permission from [13].)

### 4.2　Hydrophilicity and hydrophobicity transition

If water bonds directly to the substrate, or exchange interaction occurs between water and substrate, hydrophilicity takes place, which is subject to conditions of crystal growth – lattice matching [17] and chemical conditions [18], for instances. When encapsulated in hydrophilic nanopores [19, 20], or when wetted in hydrophilic topological configurations [21], water molecules perform in an opposite way and melt at temperatures below the bulk $T_m$.

Figure 6 shows that altering the $H_2O/SiO_2$ interface from hydrophobic to hydrophilic by water vapor plasma sputtering raises the interface shear viscosity and reduces the $\omega_H$ of the skin from the skin characteristic 3450 cm$^{-1}$ to the bulk frequency of 3200 cm$^{-1}$ [22]. Water maintains its high lubricity under the normal pressure of 1.7 MPa at pulling when confined between silica plates. However, the lubricity drops, or viscosity increases, at 0.4 MPa pressure when the polarization skin is removed by plasma sputtering [22]. Plasma sputtering removes the polarized electrons of weak binding. These observations indicate that the interface between silica and water is indeed hydrophobic but plasma sputtering alter it. Aging of the sputtered silica may recover its skin dipoles and the hydrophobicity of the silica.

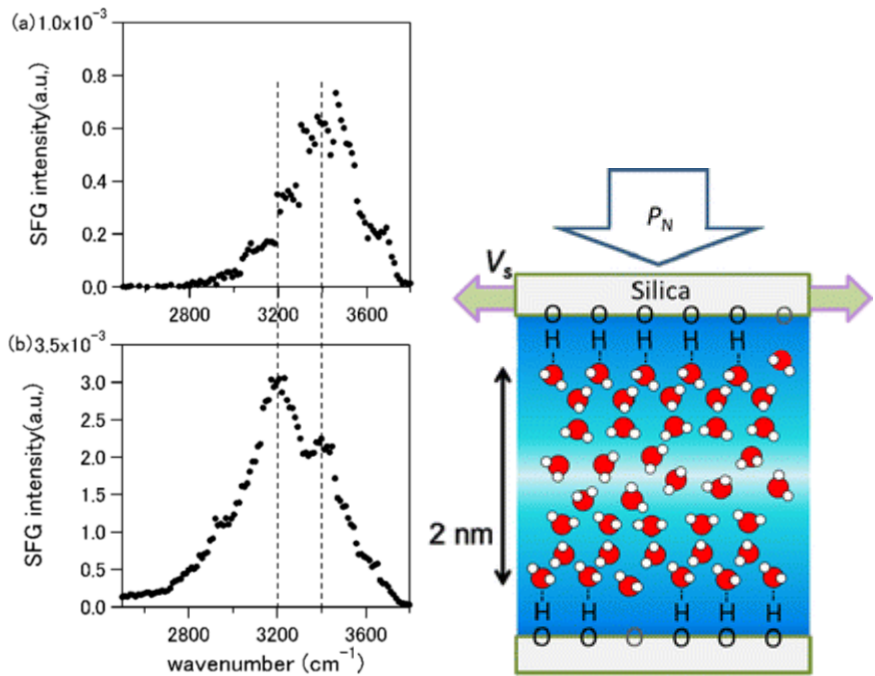

Figure 6. SFG vibrational spectra obtained from water on free (a) untreated and (b) plasma-treated silica surfaces. Phonon frequency at 3200 cm$^{-1}$ characterizes the bulk and the 3450 cm$^{-1}$ the skin. Side view illustrates the Water/Silica interface shear stress measurements under normal pressure $P_N$. (Reprinted with permission from [22].)

The presence of an air gap [11] between water and the hydrophobic substrate indicates presence of repelling between like charges on the counterparts and the elasticity of both [2, 3]. The polarization of molecules caused by both under-coordination and inter-electron-pair repulsion enhances the skin elasticity of water. The high elasticity and the high density of surface dipoles form the essential conditions for the hydrophobicity of a contacting interface [23][9].

4.3    Capillary curvature enhanced skin thermal stability

Water droplets encapsulated in hydrophobic nanopores [24] and point defects [25, 26] are thermally even more stable than the bulk water even because of the undercoordinated molecules in the curved skin. Sum frequency generation spectroscopy revealed that the skin of two adjacent molecular layers are highly ordered at the hydrophobic contacts compared with those at the flat water-air interface [27]. MD simulations suggested that freezing preferentially starts in the subsurface of water instead of the outermost layer, which remains ordered during freezing [25]. The subsurface accommodates better than the bulk the increase of volume connected with freezing. Furthermore, bulk melting is mediated by the formation of topological defects which preserve the coordination of the tetrahedral network. Such defect clusters form a defective region involving about 50 molecules with a surprisingly long lifetime [26]. These observations verify the BOLS-NEP expectations that the undercoordinated water molecules are indeed thermally stable. Therefore, a liquid layer never forms on ice [25] or surrounding defects [26].

Figure 7 shows the contact angle dependence of transition from the initial contact angles of droplet on different substrates [28]. Droplet of initially highly curved skin needs higher temperature to spead over the substrates. Likewise, a water droplet on a roughened more hydrophobic Ag skin (with nanocolumnar structures) having a greater contact angle and higher curvature, freezes 68.4 s later than on a smooth Ag surface at -4°C [29]. The formation of the proxy tip due to volume expansion at the top of the droplet indicates frozen that proceeds from the bottom of the droplet. These observations indicate that molecules at the curved skin are thermally more stable than those at the flat skin.

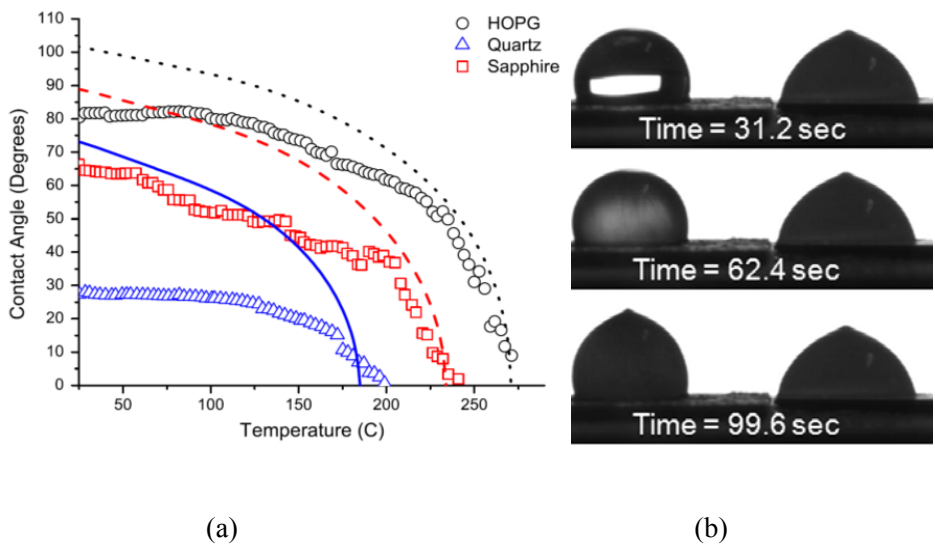

(a)            (b)

Figure 7. (a) Evolution of water contact angle on quartz, sapphire and graphite as a function of temperature (°C). (b) Water droplet on rough (left) Ag skin freezes 68.4 s later than that on the smooth Ag. The proxy tip indicates frozen. (Reprinted with permission from [28, 29].)

4.4    Bond length – phonon frequency - O 1s shift correlation

As discussed in the previous sections, the skin H-O length contracts from the bulk value of 1.0 to 0.95 Å and the O:H lengthesn from 1.68 to 1.90 Å for 200 K ice skin [30] with 6.4% elngation of O-O distance, which agrees with 5.9% measured from water skin at room temperature, expanding by In comparison, the skin O-O distance of liquid methanol contracts by 4.6% [31], which differentiates the surface tension of 72 mN/m for water from 22 mN/m for methanol. The O-O distance in the bulk varies from 2.70 [32] to 2.85 Å [33], depending on experimental conditions. Besides, the volume of water confined in 5.1 and 2.8 nm sized $TiO_2$ pores expands by 4.0% and 7.5%, respectively, with respect to the bulk water [34]. MD calculations also reveal that the $d_H$ contracts from 0.9732 Å at the center to 0.9659 Å at the skin of a free-standing water droplet containing 1000 molecules [35]. The O-O elongation results in a density loss in the water skin to a value down to 0.4 g·cm$^{-3}$ (corresponding to $d_{O-O}$ = 3.66 Å) [36, 37].

Following the same trend as 'normal' materials, molecular undercoordination imparts to water local charge densification [38-43], binding energy entrapment [39, 44-46], and nonbonding electron polarization [41]. For instance, the O 1s level shifts more deeply from the bulk value of 536.6 eV to 538.1 eV and 539.7 eV when bulk water is transformed into skin or into gaseous molecules [47, 48]. The H-O bond energy is 3.97 eV for bulk water [49] and it is 4.52 and 5.10 eV for the skin and for the gaseous monomers [50].

DFT calculation derived Mulliken charge accumulation at the skin and in the bulk of water. O increases its net charge from the bulk value of -0.616 to -0.652 e for the skin. The net charge of a water molecule increases from the bulk value of 0.022 to -0.024 e at the skin. The densification and entrapment of bonding electrons polarize the nonbonding electrons. As it has been discovered using an ultra-fast liquid jet vacuum ultra-violet photoelectron spectroscopy [41], the bound energy for an nonbonding electron in solution changes from a bulk value of 3.3 eV to1.6 eV at the water skin. The bound energy of nonbonding electrons, as a proxy of work function and surface polarization, decreases further with molecule cluster size.

Water molecular undercoordination stiffens the stiffer $\omega_H$ significantly [51, 52]. The $\omega_H$ has a peak centered at 3200 cm$^{-1}$ for bulk water, and at 3450 cm$^{-1}$ for the skins of both water and ice (see Figure 2 inset) [53]. The $\omega_H$ for gaseous molecules is around 3650 cm$^{-1}$ [54-57]. The $\omega_H$ shifts from 3200 to 3650 cm$^{-1}$ when the N of the (H$_2$O)$_N$ cluster drops from 6 to 1 [54, 58, 59]. The high frequency at approximately 3700 cm$^{-1}$ corresponds to the vibration of the dangling H-O bond radicals, with possible charge transportation in the skin of water and ice [60, 61]. DFT-MD derived that the $\omega_H$ shifts from ~3250 cm$^{-1}$ at a 7 Å depth to ~3500 cm$^{-1}$ of the 2 Å skin of liquid water [62].

Table 1 summarizes experimental information of the bond length, phonon frequency, and bond energy of water and ice under different coordination environments. With the known O-H and H:O bond length relaxation and the tetrahedrally-coordinated structure [63], we obtained the size $d_H$, separation $d_{OO}$, and mass density $\rho$ of molecules packing in water ice in the following relationships with the $d_{H0}$ and the $d_{L0}$ being the references at 4°C [49],

$$\begin{cases} d_{OO} = 2.6950\rho^{-1/3} & (Molecular\ separation) \\ \dfrac{d_L}{d_{L0}} = \dfrac{2}{1+exp\left[(d_H - d_{H0})/0.2428\right]}; & (d_{H0} = 1.0004\ and\ d_{L0} = 1.6946\ at\ 4\ °C) \end{cases}$$

(1)

With the measured $d_{OO}$ of 2.965 Å [31] as input, this relation yields the segmental lengths of $d_H$ = 0.8406 Å and $d_L$ = 2.1126 Å, which turns out a 0.75 g·cm$^{-3}$ skin mass density, much lower than the bulk value of 0.92 g·cm$^{-3}$ for ice. Indeed, the mass density of both skins suffers loss due to molecular undercoordination. Table 1 lists the $d_{OO}$, $d_x$, $\rho$, and $\omega_x$ for the skin and the bulk of water and ice in comparison to those of ice at 80 K and water dimers with the referenced data as input.

Table 1 Experimentally-derived skin supersolidity ($\omega_x$, $d_x$, $\rho$) of water and ice.

|  | Water (298 K) | | Ice (253K) | Ice(80 K) | Vapor |
| --- | --- | --- | --- | --- | --- |
|  | bulk | skin | bulk | Bulk | dimer |
| $\omega_H$(cm$^{-1}$) | 3200[53] | 3450[53] | 3125[53] | 3090[64] | 3650[55] |
| $\omega_L$(cm$^{-1}$)[64] | 220 | ~180[63] | 210 | 235 | 0 |
| $d_{OO}$(Å) [49] | 2.700[32] | 2.965[31] | 2.771 | 2.751 | 2.980[31] |
| $d_H$(Å) [49] | 0.9981 | 0.8406 | 0.9676 | 0.9771 | 0.8030 |
| $d_L$(Å) [49] | 1.6969 | 2.1126 | 1.8034 | 1.7739 | ≥2.177 |
| $\rho$(g·cm$^{-3}$) [49] | 0.9945 | 0.7509 | 0.92[65] | 0.94[65] | ≤0.7396 |

## 4.5  Viscoelasticity, repulsion, and hydrophobicity

The polarization of molecules enhances the skin repulsion and viscoelasticity. The high viscoelasticity and the high density of skin dipoles are essential to the hydrophobicity and lubricity at contacts [23]. According to the BOLS-NEP notion, the local energy densification stiffens the skin and the densely and tightly entrapped bonding charges polarize nonbonding electrons to form anchored skin dipoles [66].

Reducing the number of molecular layers of the skin increases local surface tension $\gamma$, and viscosity $\eta_s$ and $\eta_v$ [30]. The O:H-O cooperative relaxation and associated electron entrapment and polarization enhances the surface tension from 31.5 for 15 layers to 73.6 mN/m for five layers, which approaches the measured value of 72 mN/m for water skin at 25°C. The skin viscosity increases from 0.07 to 0.019 $10^{-2}$mN·s/m$^2$. The bulk $\eta_v$ changes insignificantly from 0.027 to 0.032 for five layer thick skin. Generally, the viscosity of water reaches its maximum at a temperature around the $T_m$ [67].

The negative charge gain and the nonbonding electron polarization provide electrostatic repulsive force that not only lubricates ice but also the hydrophobicity of water skin. Measurements of an elastic modulus of 6.7 GPa have verified the presence of the repulsive forces between a hydrated mica substrate and the tungsten contacts at 24°C under 20 – 45% relative humidity (RH) [68]. Monolayer ice also forms on a graphite surface at 25% RH and 25°C [69]. These observations and the present numerical derivatives evidence the presence of the supersolidity with repulsive forces because of bonding charge densification, surface polarization and $T_m$ elevation.

## 4.6  Thermal relaxation of skin tension

Instead of the energy loss upon surface formation-conventionally called surface energy, the energy gain of a unit volume or the cohesive energy gain of a discrete atom in the skin of certain thickness due to atomic undercoordination governs the performance of a surface [70]. Energy density determines the local elasticity and yield strength according to their dimensionality [Pa = Force/Area = energy/volume], $Y_z \propto E_z/d_z^3$ and the cohesive energy the thermal stability $T_{Cz} \propto zE_z$. The temperature dependence of skin energy density, $\gamma_{di}$, and the skin atomic coherency, $\gamma_{fi}$, follows the relationship [70]:

$$\gamma_{di}(T) \propto \frac{E_i(T)}{d_i^3(T)} = \frac{E_i(0) - \int_0^T \eta_{li}(t)dt}{d_i^3 \left(1 + \int_0^T \alpha_i(t)dt\right)^3}$$

Where $\eta_1$ is the specific heat in Debye approximation and $\alpha$ is the thermal expansion coefficient. The relative change of the skin energy density

$$\frac{\gamma_{di}(T)}{\gamma_d(0)} \cong \frac{E_i(0) - \int_0^{T/\Theta_D} \eta_{1i}(t/\Theta_D)dt}{E_b(0)}$$

$$\approx \exp\left(-3\int_0^T \alpha_i(t)dt\right) \begin{cases} 1 - \dfrac{\int_0^T \eta_1(t/\Theta_D)dt}{E_b(0)}, & (T \leq \theta_D) \\ 1 - \dfrac{\eta_1 T}{E_b(0)}, & (T > \theta_D) \end{cases}$$

Figure 8 shows the reproduction of the measured temperature dependence of surface tension of liquid H$_2$O, which turns out the Debye temperature $\Theta_D$ and molecular cohesive energy $E_b(0)$ with the thermal expansion coefficient as input. One molecule connects to the surrounding by four identical nonbonds whose energy is estimated as $E_L$= 0.38/4 = 0.095 eV, in the 0.1 eV order. The skin O:H elongation lowers the $E_L$ and down to 0.036 eV for a dimer [6].

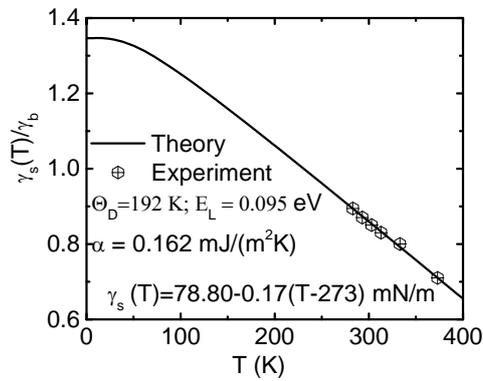

Figure 8. Numerical reproduction of the surface tension turns out the Debye temperature and the molecular cohesive energy in water skin [70].

4.7    Water skin supersolidity

It is convenient to adapt the concept of supersolidity from the superfluidity of solid $^4$He at mK temperatures. Atomic undercoordination-induced local strain and the associated quantum entrapment and polarization rationalize $^4$He superfluidity and supersolidity - elastic and repulsive between locked dipoles at fragment contacts. The skins of $^4$He fragments are highly elastic and frictionless with repulsion between them when in motion [66].

As justified above, the skins of water and ice form an extraordinary supersolid phase that is elastic [53], hydrophobic [9, 12], polarized [41, 60] and thermally stable [71], with densely entrapped bonding electrons [44, 47, 48, 72] and ultra-low-density [31]. The fewer the molecular neighbors there are, the smaller the water molecule size is, the greater the molecular separation is, and therefore the greater the supersolidity will be. The supersolid skin is responsible not only for the slipperiness of ice but also for the hydrophobicity and toughness of water skin.

## 5  Insight extension: Hydrophobicity and hydrophilicity

### 5.1  Wetting

#### 5.1.1  Definition: Young equation

Wetting is the process of making contact between a solid and a liquid. Controlling the wettability of solid materials is a classical and key issue in surface engineering. Typical examples for wetting-dependent processes in daily life, biology and industry include adhesion, cleaning, lubricating, painting, printing, and many more.

Most processes that involve liquids deal with situations where the free surface of the liquid meets a solid boundary, thus forming the so-called three-phase-contact line - solid-liquid-gas. The contact line can move along the solid surface, leading to "wetting" or "dewetting". The interaction between a liquid and a solid involves three interfaces; the solid-liquid interface, the liquid-vapor interface and the solid-vapor interface.

Each of these interfaces has an associated surface tension, $\gamma$, which represents the energy required to create a unit area of that particular interface. A different approach is to regard $\gamma$ as a force acting on the water drop. This approach is shown in Figure 9, where $\gamma$ appears as an arrow. At equilibrium, force equilibrium along the X axis provides a relation between the angle, $\theta$, and the surface tensions of the three interfaces. Young equation formulates the surface wetting from the perspective of force equilibrium, as illustrated in Figure 9:

$$\cos\theta = \frac{\gamma_{SG} - \gamma_{SL}}{\gamma_{LG}}$$

Where $\gamma_{SG}$, $\gamma_{SL}$ and $\gamma_{LG}$ are the surface tensions of interfaces solid/gas, solid/liquid and liquid/gas respectively. θ is the angle between a liquid drop and a solid surface, called the contact angle.

The magnitude of Young's contact angle is the result of energy minimization. If the liquid-gas surface tension is smaller than the solid-gas surface tension ($\gamma_{LG} < \gamma_{SG}$), the liquid-solid interface will increase to minimize energy. As the drop wets the surface, the contact angle approaches zero, leading to complete wetting. Other ratios of $\gamma_{LG}$ and $\gamma_{SG}$ will lead to the formation of drops of different shapes. A hydrophilic surface is defined as a surface where 0° < θ < 90°, and hydrophobic surface is a surface where θ ≥ 90°. The surface tensions of different substances in contact with the gas phase vary over a wide range.

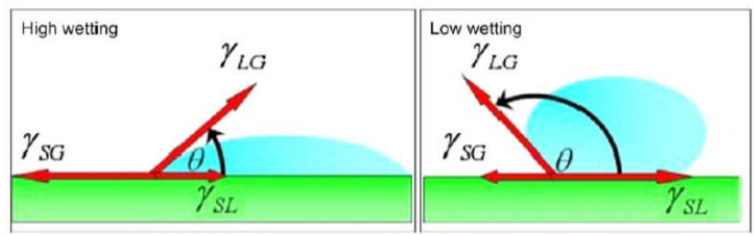

Figure 9. Illustration of Young equation for the surface energy. On the left, there is much wetting and the contact angle is small. On the right little wetting and the contact angle is large. The drawing derives Young force equations, $\cos\theta = (\gamma_{SG} - \gamma_{SL})/\gamma_{LG}$ [73].

### 5.1.2 Interaction between surfaces and liquids

In order to understand why different surfaces and liquids form different contact angle, one must consider the interaction between the liquid and the surface and between the liquid molecules themselves. The adhesive force between the surface and the liquid causes the drop to spread and wet the surface, and the cohesive force within the liquid drop causes it to ball up and avoid contact with the surface. For example, let's consider the contact angle water and three types of surfaces - polymer, metal and oxide.

- Some polymers, such as PVC and Teflon, form mainly Van der Waals bonds with a water drop placed on them. These bonds are weak relative to the hydrogen bonds within the drop, so that the water prefers to bond with itself and not with the surface. The result is a bead shaped drop that almost doesn't wet the surface. These are hydrophobic surfaces.
- Oxide surfaces can form hydrogen bonds with the water in the drop. These are strong bonds, and so the water prefers to wet the surface rather than ball into a bead. These are hydrophilic surfaces.
- Metal surfaces don't form strong hydrogen bonds with water, however their polarity is greater than that of PVC or Teflon, so their affinity to water is greater, the result is that their contact angle is usually between that of the previous surfaces.

### 5.1.3 Controlling the contact angle

The surface tension and contact angle can be controlled through different methods.

1. Temperature: generally, surface tension decreases as the temperature increases. The dependence of surface tension on temperature is usually approximately linear reaching zero at the critical temperature, as in Figure 1b.
2. Surface roughness: if the surface is rough and water penetrates into the grooves, whatever trend the flat surface showed will be enhanced: a hydrophobic surface will become even more hydrophobic and a hydrophilic surface will become more hydrophilic. This is Wenzel's notion. According to Cassie and Baxter's notion, the water drop doesn't penetrate into the surface grooves and instead lies on top of them, so that air bubbles are trapped inside. The trapped air increases the contact angle and the surface becomes superhydrophobic. Figure 10 illustrates both models.
3. Electro-wetting: when a potential difference is applied between the liquid and the solid surface, the electric force at the corners of the drop pulls it down onto the surface, lowering the contact angle.
4. Chemical modifications: chemical modification of the surface can lead to a change in its surface tension. For example, adding polar, hydrophilic groups to the surface will lead to a lower contact angle.

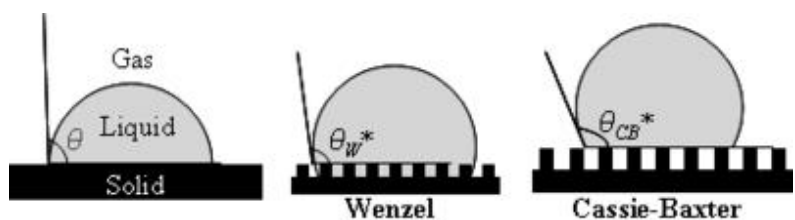

Figure 10. Illustration of the skin roughness dependence of the contact angle. (a) Smooth skin and rough skins with (b) water penetration (Wenzel's model) and (c) gas bubbles entrapment (Cassie-Baxter model) grooves.

## 5.2 Super hydrophobicity and superhydrophilicity

### 5.2.1 Superhydrophobicity

The phenomena of superhydrophobicity, superfluidity, superlubricity and supersolidity (4S) at the nanometer-sized contacts of liquid-solid or solid-solid share the common characteristics of chemically non-stick, mechanically elastic, and kinetically frictionless in motion. Although the 4S occurrences have been extensively investigated, mechanism behind the common characteristics remains in it infancy. The BOLS-NEP notion provides an energetic and electronic mechanism indicating that Coulomb repulsion between "dipoles pinned in the elastic skins or the supersolid covering sheets of liquid droplets" dictates the 4S [66].

The localized energy densification makes the skin stiffer and the densely- and tightly-trapped bonding charges polarize nonbonding electrons, if exist, to form locked skin dipoles. In addition, the sp-orbit hybridization of F, O, N, or C upon reacting with solid atoms generates nonbonding lone pairs or unpaired edge electrons that induce dipoles directing into the open end of a surface. Such a Coulomb repulsion between the negatively charged skins of the contacting objects not only lowers the effective contacting force and the friction but also prevents charge from being exchanged between the counterparts of the contact. Being similar to magnetic levitation, such Coulomb repulsion provides the force driving the 4S.

### 5.2.2 Wenzel-Cassie-Baxter models

The following theories describe the underlying mechanism for the 4S phenomena:

1) Young's theory in terms of surface tension and interface energies [74].
2) Wenzel and Cassie-Baxters' law [75, 76] of surface roughness for superhydrophobicity.
3) Electrical double layer (EDL) scheme for the superfluidity [77].
4) Prandtl–Tomlinson (PT) theory [78, 79] of the superposition of the slope of atomic potential and multiple-contact effects [80] for atomic scale quantum friction.

Wenzel's law suggested that if the surface is rough and water penetrates into the grooves, whatever trend the flat surface showed will be enhanced: a hydrophobic surface will become even more hydrophobic and a hydrophilic surface will become more hydrophilic. According to Cassie and Baxter's notion, the water drop doesn't penetrate into the surface grooves and instead lies on top of them, so that air bubbles are trapped inside. The trapped air increases the contact angle and the surface becomes superhydrophobic.

Many of these superhydrophobic materials found in nature display characteristics fulfilling Wenzel-Cassie-Baxters' law [81] stating that the surface contact angle can be increased by simply roughing up the surface, i.e., the surface roughness and the contact area are suggested to be the factors of dominance. For instances, fluids can slip frictionlessly past pockets of air between textured surfaces with micrometer-scale grooves or posts of tiny distances [82]. The slip length for water is almost ten times longer than previously achieved, indicating that engineered surfaces can significantly reduce drag in fluid systems. On the base of Cassie-Baxters' law and thermodynamics considerations, Fang et al [83] and Li et al [84] designed tunable superhydrophobic surfaces to control the directional motion of water droplets by varying the pillar width and spacing simultaneously. Varying the gradient of the stiffness of a micro-beam could also drive directional movement of liquid droplets on a microbeam [85].

A water strider statically standing on water can bear a load up to ten times its body weight with its middle and hind legs, which tread deep puddles without piercing the water surface [1]. This fact illustrates the superhydrophobicity of the water strider legs due to "bio-wax" coatings. Another comparative experiment [86] using the real water strider legs and artificial legs made of wax-coated steel wires revealed that the adaptive-deformation capacity of the real leg through its three joints makes a more important contribution to the superior load-bearing ability than the superhydrophobicity.

BOLS-NEP notion combines both the Wenzel and Cassie-Baxter models from skin polarization point of view. Atomic undercoordination becomes more pronounced when the curvature of the proxy is increased, which enhances the entrapment –polarization defined by the BOLS-NEP notion [16] and then the Wenzel effect. Air pockets will form underneath water droplets if the rough skin is hydrophobic. If the hydrophilic skin is roughens, it will be even hydrophilic and no air pockets will present. Atomic undercoordination induced local entrapment is global yet the polarization is subjective, which is why Wenzel's model works. Pt, Co and graphite skin show entrapment dominance while Cu, Ag, Au, Rh, W, Mo, and graphite point defects demonstrate polarization dominance; most oxides, nitrides, fluorides skins and defects are polarization dominance because of nonbonding electron pairs [87].

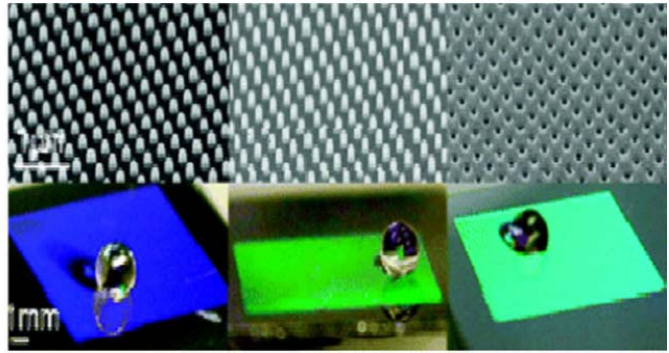

Figure 11. Substrate patterns and the superhydrophobicity of water droplets [88]. Atomic undercoordination at the sharp tip proxy undergoes bonding electron quantum entrapment and densification and nonbonding electron polarization, which results in the high elasticity and strongly polarized skin, being responsible for the superhydrophobicity of the substrate [66].

### 5.2.3 BOLS-NEP notion

The 4S occurrences result from the reduction of the friction force ($f_r = \mu N$ with $\mu$ being the friction coefficient and N the contacting force). The lowering of the $f_r$ will reduce the process of friction or the extent of phonon and electron excitation. One surprising fact is that these 4S effects share a general identity of non-sticky and frictionless motion-with lowered effective contacting pressure and reduced friction coefficient. Skins for both the water and the substrate must be hydrophobic to ensure the superhydrophobicity at working.

The 4S phenomena must share a common elastic and repulsive origin in addition to the energetic and geometric descriptions of the existing models. Considerations from the perspectives of surface roughness, air pocket, and surface energy seem insufficient because the chemistry and the charge identities do alter at the surface skin up to two interatomic spacings [89]. In particular, the hydrophobicity-hydrophilicity recycling effect caused by UV irradiation and the subsequent dark aging is beyond the scope of Cassie's law and the PT mechanism of air pockets dominance. Furthermore, the superhydrophobicity of alkanes, oils, fats, wax, and the greasy and organic substances is independent of the surface roughness.

### 5.2.4 Super hydrophobicity-hydrophilicity transition

Superhydrophobic materials have surfaces that are extremely difficult to wet, with water contact angles in excess of 150° or even greater, see Figure 11. Surfaces with ultra-hydrophobicity have aroused much interest with their potential applications in self-cleaning coatings, microfluidics, and biocompatible materials and so on. Many physical chemical processes, such as adsorption, lubrication, adhesion, dispersion, friction, etc., are closely related to the wettability of materials surfaces [90, 91]. Examples of hydrophobic molecules include alkanes, oils, fats, wax, and greasy and organic substances with C, N, O, or F as the key constituent element.

What is even more amazing is that the hydrophobic surface can switch reversibly between superhydrophobicity and superhydrophilicity when the solid surface is subject to UV radiation [92] which results in electron-hole pair creation [93]. After being stored in the dark over an extended period, the hydrophilicity is once again lost. The dipoles can be demolished by UV radiation, thermal excitation, or excessively applied compression due to ionization or sp orbit de-hybridization.

The UV radiation with excitation energy around 3.0 eV could break chemical bonds and ionize surface atoms, which could turn the hydrophobic surface to be hydrophilic, as it has widely been observed. $Ar^+$ sputtering the surface is expected to have the same function of removing dipole or monopole temporarily. If the polarized electrons were removed by UV irradiation, sputtering, or thermal excitation, the 4S characteristics would be lost. Aging of the specimen will recover the surface charges. The effect of UV radiation reversing effect is the same as that observed in the surface magnetism of noble metal clusters and the dilute magnetism of oxide nanostructures [94-96]. Thermal annealing at temperatures of 600 K or above, oxygen orbital de-hybridization takes place and the lone pair induced Cu surface dipoles vanish [97]. However, aging the samples in the ambient will recover the sp-hybridization and the dipoles as well. Surface bias to a certain extent may also cause the depletion of the locked charges though this expectation is subject to verification. Overloaded pressure in the dry sliding will overcome the Coulomb repulsion, as the energy dissipation by phonon and electron excitation could occur under the applied pressure. On the other hand, a sufficiently large difference in the electro-affinity between the contact media, chemical bond may form under a certain conditions such as heating, pressure, or electric field, the interface will be adherent.

## 5.3    Superfluidity in nanochannels

This understanding may extend to the superfluidity of $^4$He [66] and water droplet flowing in carbon nanotubes [98]. It is understandable now why the rate of the pressure-driven water flow through carbon nanotubes is orders higher in magnitude and faster than is predicted from conventional fluid-flow theory [99]. It is within expectation that the narrower the channel diameter is, the faster the flow of the fluid will be [98, 100], because of the curvature-enhanced supersolidity of the water droplet interacting with hydrophobic carbon nanotubes.

The transport of fluid in and around nanometer-sized objects with at least one characteristic dimension below 100 nm enables the superfluidic occurrence that is impossible on bigger length scales [101]. Nanofluids have significantly greater thermal and mass conductivity in nanochannels compared with their base fluids [102]. The difference between the nanofluid and the base fluid is the high value of surface-to-volume ratio that increases with the miniaturization of the dimensions of both the fluid and the channel cavity in which the fluid is flowing. This high ratio in nanochannels results in surface-charge-governed transport, which allows ion separation and is described by an electrokinetic theory of electrical double layer (EDL) scheme [77]. The EDL channel can be operated as field-effect transistors to detect chemical and biological species label-free, and transport through nanochannels leads to analyte separation and new phenomena when the EDL thickness becomes comparable to the smallest channel opening.

On the other hand, the rate of the pressure-driven water flow through carbon nanotubes (CNTs) is orders higher than predictions by conventional fluid-flow theory [99]. The thinner the channel cavity is, the faster the fluid-flow rate will be under the same pressure [98]. This high fluid velocity results from an almost frictionless interface between the CNT wall and the fluid droplets [103, 104]. A MD calculation [105] suggested that water flow in CNT could generate a constant voltage difference of several mV between the two ends of a CNT, due to interactions between the water dipole chains and charge carriers in the CNT, which might also contribute to the abnormal frictionless fluid flow in the CNT.

Although the crystal defects have been recognized as the key to the supersolidity of $^4$He solid, correlation between the defects and the superelasticity and superfluidity is yet to be established. Therefore, a deeper insight into the chemical nature of the surfaces is necessary for one to gain a consistent understanding of the origin for the 4S.

### 5.4 BOLS-NEP formulation

Figure 12a shows the theoretically predicted curvature ($K^{-1}$) dependence of the skin charge density, elasticity (energy density), and potential trap depth (core level shift) of the outermost shell of a spherical dot [106]. The volume average correspond to the size dependence of the elastic modulus such as ZnO [70] and the core-level shift of nanostructures [107].

As illustrated in Figure 12b, the drop and the wall surface are likely charged (green dots) and repelling each other, which ensures water skin not only high elasticity but also electric repulsive – under compression instead of tension. The droplet will lose its viscosity and becomes frictionless unless the surface dipoles are removed. Such a system runs in a way more like a "maglev train".

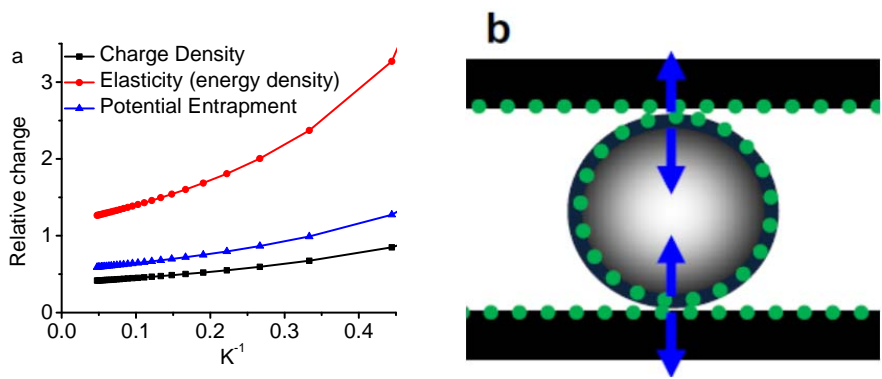

Figure 12. (a) Curvature ($K^{-1}$) dependence of the skin charge density, elasticity (energy density), and potential trap depth of the outermost shell of a spherical dot. (b) A water droplet of supersolid skin flowing through a nanochannel is subject to electro repulsion. The smaller the droplet is, the higher the supersolidity extent will be [106].

The superhydrophobicity phenomenon can be explained from the viewpoints of surface chemistry, energy and charge density enhancement. If the air pockets beneath a droplet on a sinusoidal substrate are open to the atmosphere, the superhydrophobic state can exist only when the substrate is hydrophobic, and that the geometric parameters of the microstructure have a great influence on the wetting behavior. Being similar to the superfluidity, polarization of the surface or the presence of lone pair electrons happens to both the fluidic drop and the material. The charged surface repels the ambient charged particles, such as water molecules, to result in superhydrophobicity. The UV radiation removes the polarized charges and the dark storage recovers the surface dipoles, being the same as the surface magnetism of noble metal clusters and the dilute magnetism of oxide nanostructures [95, 96].

Increasing pressure normally promotes a 'normal' liquid freezing, shifting the melting point to higher temperatures [108]. The melting temperature is proportional to the atomic cohesive energy, $\Delta T_m \propto \Delta(zE_z) \propto -P\Delta V/N$. z is the atomic coordination number and $E_z$ the cohesive energy per bond. V is the volume and N the total number of atoms of the substance at question.

In water it is the opposite [109], ice will melt when subjected to pressure (at least until 2100 atmospheres when water freezes at -22°C). Transposing this explanation into ice skating, it was thought that ice melted under the skate's pressure producing the lubricating layer of water of at most -3.5 °C. However, this mechanism does not explain why it is still possible to skate at temperatures below -3.5 °C, as the pressures exerted by the skates are not sufficiently high to melt the ice at lower temperatures. The optimum temperature for ice hockey is -9 °C and it is still possible to ski and skate at temperatures as low as -35°C. Nevertheless, why pressure lowers the melting pint of ice instead of raising it remained unclear till 2012 when Sun et al [109] clarified that compression lengthens and softens the H-O bond whose cohesive energy loss dominates the $T_m$ drop of ice.

6       Summary

Undercoordination-induced O:H-O bond relaxation and the associated binding electron entrapment and the nonbonding electron dual polarization clarify the anomalous behaviour of water molecules with fewer than four nearest neighbours - in particular, the skin supersolidity of water and ice. Agreement between numerical calculations and experimental observations verified the following:

1) Undercoordination-induced O:H-O relaxation results in the supersolid phase that is elastic, hydrophobic, thermally more stable, and less dense, which dictates the unusual behaviour of water molecules at the boundary of the O:H-O networks or in the nanoscale droplet.
2) H-O bond contraction densifies and entraps the core and bonding electrons; H-O bond stiffening shifts positively the O1s energy, the $\omega_H$ and the $T_m$ of molecular clusters, surface skins, and ultrathin films of water.
3) The dual polarization makes the skins hydrophobic, viscoelastic, and frictionless.

4) Neither a liquid skin forms on ice nor a solid skin covers water; rather, a common supersolid skin covers both. The supersolid skin causes slippery ice and toughens water skin.